\documentclass[journal=nalefd, manuscript=letter]{achemso}

\usepackage{graphicx}
\usepackage{dcolumn}
\usepackage{bm}
\usepackage{braket}
\usepackage{verbatim}
\usepackage{lipsum}
\usepackage{amsmath,amssymb}

\usepackage[dvipsnames]{xcolor}

\usepackage{enumitem}
\usepackage{makecell}
\usepackage{relsize}
\usepackage{hyperref}

\newcommand{\beq}{\begin{equation}}
\newcommand{\eeq}{\end{equation}}
\newcommand{\beqarray}{\begin{eqnarray}}
\newcommand{\eeqarray}{\end{eqnarray}}

\title{Fast, accurate, and error-resilient variational quantum noise spectroscopy}
\author{Nanako Shitara}
\affiliation{Department of Chemistry, University of Colorado Boulder, Colorado 80309, USA}
\alsoaffiliation{Department of Physics, University of Colorado Boulder, Colorado 80309, USA}
\author{Andr\'es Montoya-Castillo}
\affiliation{Department of Chemistry, University of Colorado Boulder, Colorado 80309, USA}
\email{Andres.MontoyaCastillo@colorado.edu}
\date{\today}

\begin{document}

\begin{abstract}
Detecting and characterizing decoherence-inducing noise sources is critical for developing robust quantum technologies and deploying quantum sensors operating at molecular scales. However, current noise spectroscopies rely on severe approximations that sacrifice accuracy and precision. We propose a novel approach to overcome these limitations. It self-consistently extracts noise spectra that characterize the interactions between a quantum sensor and its environment from commonly performed dynamical decoupling-based coherence measurements. Our approach adopts minimal assumptions and is resilient to measurement errors. We quantify confidence intervals and sensitivity measures to identify experiments that improve spectral reconstruction. We employ our method to reconstruct the noise spectrum of a nitrogen-vacancy sensor in diamond, resolving previously undetected nuclear species at the diamond surface and revealing that previous measurements had overestimated the strength of low-frequency noise by an order of magnitude. Our method uncovers previously hidden structure with unprecedented accuracy, setting the stage for precision noise spectroscopy-based quantum metrology. 

\end{abstract}

\maketitle


Characterizing and controlling decoherence-inducing environmental noise is key to developing next-generation quantum processors, memories, and sensors. Dynamical decoupling (DD) sequences are widely used to extend coherence lifetimes necessary for processors and memories~\cite{Hahn1950, Carr1954, Meiboom1958, Viola1998, Viola1999, Vitali1999, deLange2010, Du2009, Moisanu2024}, and underlie popular quantum sensing protocols~\cite{Jiang2023, Jerger2023, Machado2023, Muller2014, Ajoy2015, Du2024, Degen2017}. By encoding power spectra that quantify environmental interactions and timescales, the dynamics generated by these DD sequences can reveal fluctuations in magnetic~\cite{Zhao2012} and electric fields~\cite{Dolde2011} and mechanical forces~\cite{Ovartchaiyapong2014}, and have been used to characterize protein~\cite{Shi2015, Lovchinsky2016} and DNA~\cite{Shi2018} motions, topological defects~\cite{Jenkins2019, Dovzhenko2018}, and nanoscale structure~\cite{Smits2019, Abobeih2019}. Indeed, access to these power spectra can guide optimal decoherence mitigation through filter design principles~\cite{Biercuk2011} and facilitate physically insightful quantum sensing~\cite{Alvarez2011, Bylander2011, Soetbeer2021}. However, while many sophisticated noise spectroscopies aim to extract accurate and precise noise spectra from accessible coherence measurements,~\cite{Chalermpusitarak2021,Norris2016, Soetbeer2021, Sun2022,Lupke2020,Tripathi2024, Wang2024, Sung2021, Wudarski2023, Rovny2022, PazSilva2019, Khan2024, Sangtawesin2019, Monge2023, Chrostoski2022, Myers2014, Dial2013, Barr2024, Malinowski2017, Ferrie2018, Frey2017, Wise2021} this remains a challenging task.

Arguably most widely used~\cite{BarGill2012, Romach2015, Biercuk2009, Connors2022, Bylander2011, Muhonen2014, Medford2012, DeOliveira2017, Kim2015, Ishizu2020}, \'Alvarez and Suter's DD noise spectroscopy (DDNS)~\cite{Alvarez2011} quantifies the noise spectrum at specific frequencies by applying many effectively instantaneous $\pi$ pulses, which can be experimentally challenging. Conversely, Fourier Transform Noise Spectroscopy (FTNS)~\cite{Vezvaee2024} uses only zero or one intermediary pulses to give better resolution over wider frequency ranges, albeit requiring extensively averaged data and high temporal resolution. Thus, these methods work in opposite limits: DDNS captures the noise spectrum at \textit{local} frequency values at the cost of many highly engineered pulses, while FTNS applies only a few pulses to obtain a \textit{nonlocal} view of the noise spectrum at the cost of high statistical certainty and temporal resolution in the coherence measurements. This raises the question: can one reap the benefits of these complementary limits with a single method?


\begin{figure*}
\hspace{-.5cm}{\includegraphics[width=0.8\textwidth]{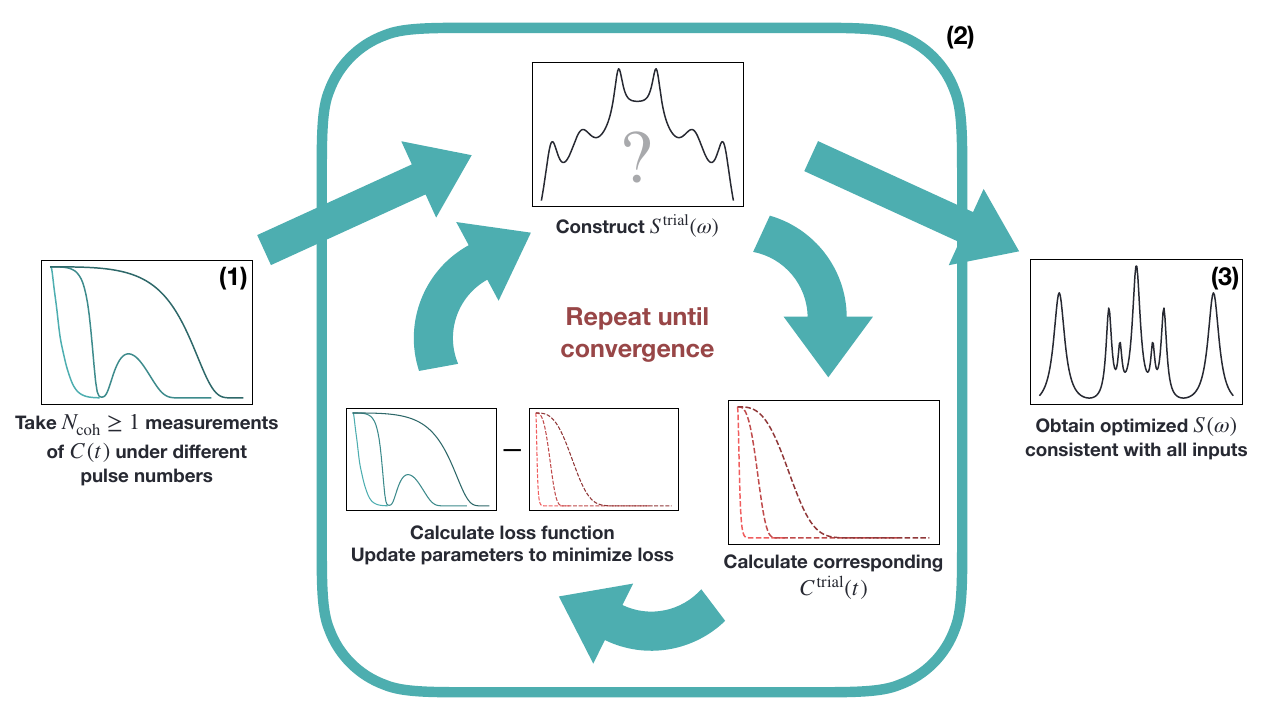}}
\vspace{-10pt}
\caption{Schematic for the {variational quantum} noise spectroscopy (VQNS) protocol. (1) Measure one or more coherence decays under FID, SE, or CPMG dynamical decoupling pulse sequences. (2) Use the measurements as input into our iterative optimization that minimizes the deviation of the measured and trial coherences obtained from a trial spectrum by modifying the coefficients of a Lorentzian basis expansion. (3) Our VQNS yields a converged trial spectrum once the iterative loop satisfies an error threshold, $\xi$, for the deviation-based loss function.} 
\label{fig-schematic}
\vspace{-10pt}
\end{figure*}


Here, we propose and demonstrate such an approach: variational quantum noise spectroscopy (VQNS). Like DDNS and FTNS, VQNS tackles the pure dephasing limit, appropriate across platforms spanning color centers~\cite{Pezzagna2021, Zhao2012PRB, Wang2012, breitweiser2024quadrupolar, zvi2025engineering, kawai2019nitrogen, wong2024coherent, monge2023spin, ermakova2013detection, kumar2024room, farfurnik2023all}, trapped ions~\cite{Biercuk2009, Harty2014}, neutral atoms~\cite{Graham2022, Wintersperger2023}, and molecular magnets~\cite{zadrozny2015, aravena2020, Soetbeer2021}. Our VQNS employs $N_{\rm coh} \geq 1$ coherence measurements subject to pulse sequences to identify a noise spectrum consistent with \textit{all} input measurements up to a specified error threshold (see Fig.~\ref{fig-schematic}). Indeed, VQNS extracts accurate noise spectra with associated confidence intervals even from underconverged experimental measurements with low temporal resolution. 
We achieve this by coupling modern optimization algorithms~\cite{Kingma2014} with a basis of analytical responses arising from the application of arbitrary pulse sequences, including Ramsey free induction decay (FID), Hahn/spin echo (SE)~\cite{Hahn1950}, and Carr-Purcell-Meiboom-Gill (CPMG)~\cite{Carr1954, Meiboom1958} pulse sequences. We combine the confidence intervals with pulse frequency sensitivities to propose tailored experiments leveraging pulse sequences that tighten confidence intervals over localized frequency ranges. Finally, we apply our VQNS to \textit{experimental} data from NV centers, where it correctly identifies hydrogen nuclei near the NV surface, which DDNS had missed, and which required orthogonal measurements to confirm in the original study~\cite{Romach2015}, illustrating the increased sensitivity VQNS offers as a quantum sensing protocol. 

Under pure dephasing, phase randomization happens much faster than population relaxation ($T_2^* \ll T_1$), and the sensor Hamiltonian becomes $\hat{H} = \frac{1}{2}[\Omega + \hat{\beta}(t)]\hat{\sigma}_z$, where $\Omega$ is the sensor frequency, and $\hat{\beta}(t)$ encapsulates the environment fluctuations that cause sensor dephasing. When $\hat{\beta}(t)$ is a stationary process obeying Gaussian statistics (i.e., $\langle \beta(t) \rangle = 0$, $D(t_1, t_2) = \langle \beta(t_1)\beta(t_2) \rangle = D(t_1 - t_2)$, and all cumulants higher than second order vanish), the coherence function $C(t) = \vert \langle \rho_{01}(t) \rangle \vert$ subject to a control pulse sequence becomes
\begin{equation}
    C(t) = e^{-\chi(t)} = \mathrm{exp}\left[ - \frac{1}{4\pi}\int_{-\infty}^\infty d\omega \, S(\omega) F(\omega t) \right].\label{eq:FFformalism}
\end{equation}
The central goal of noise spectroscopy is to access the noise power spectrum, $S(\omega)= \int_{-\infty}^\infty dt\, e^{i\omega t} \mathrm{Re}\  D(t)$\cite{Szankowski2017, Bylander2011, Degen2017, Norris2016}. The filter function, $F(\omega t)$, is the Fourier transform of the switching function that encodes the temporal distribution of $\pi$ pulses applied to the system~\cite{Szankowski2017,Uhrig2007,Uhrig2008,Cywinski2008}. Yet, while filter functions are known analytically (see Supplementary Information (SI)~Sec.~I), and one may measure $C(t)$ via techniques like NMR or EPR, extracting $S(\omega)$ under an arbitrary control pulse sequence remains a fundamental challenge. This is because analytically inverting Eq.~\ref{eq:FFformalism} for arbitrary pulse sequences to recover $S(\omega)$ is generally impossible. DDNS and FTNS provide local and nonlocal information about the noise spectrum, respectively, albeit at great cost: in the infinite pulse limit (DDNS), or the low statistical error and high temporal resolution limits (FTNS). 

Our VQNS matches the virtues of DDNS and FTNS and surpasses them: it is resilient to statistically underconverged data, accesses broad frequency ranges at high resolution, and constrains spectral reconstructions by employing data from multiple pulse sequences simultaneously. We achieve this by reformulating the noise spectrum extraction as an optimization problem that minimizes the deviation between experimental coherences arising from the true noise spectrum, $S(\omega)$, and those obtained from a trial noise spectrum, $S^{\rm trial}(\omega)$. Mathematically, one must solve $d|C^{\rm exp}(t) - C^{\rm trial}(t; \boldsymbol{\theta})|^2/d\boldsymbol{\theta} = 0$, where $\boldsymbol{\theta}$ denotes parameters that determine $S^{\rm trial}(\omega, \boldsymbol{\theta})$, which produces the trial coherence curve, $C^{\rm trial}(t; \boldsymbol{\theta})$. 

How should one pick $\boldsymbol{\theta}$? For example, $\boldsymbol{\theta}$ may be the coefficients in a functional expansion of $S^{\rm trial}(\omega)$ over a complete, orthonormal set of polynomials (e.g., Legendre or Chebyshev). While this is reasonable for many applications~\cite{Karniadakis2005, Boyd2001}, computing the trial coherence curve from a trial noise spectrum requires one to perform the integral in Eq.~\ref{eq:FFformalism}. Because this integral is generally not analytically solvable for arbitrary pulse sequences, one must resort to quadrature of highly oscillatory functions over an infinite domain, which is slow and renders the optimization process inefficient compared to the ideal case where the integrals can be performed analytically. We overcome this challenge by employing an overcomplete Lorentzian basis~\cite{Klauder1962, Meier1999, Liu2014, Dirdal2013} for which the integral in Eq.~\ref{eq:FFformalism} can be solved analytically (see SI~Sec.~I), even under arbitrary pulse sequences \cite{Szankowski2018}. We represent $S^{\rm trial}(\omega)$ as a sum of $N_{\rm basis}$ symmetrized Lorentzians,
\begin{equation}
    S^{\rm trial}(\omega) = \sum_{i=1}^{N_{\rm basis}} L_i(\omega),
\end{equation}
satisfying $S(\omega) = S(-\omega)$. Each Lorentzian, 
\begin{equation}
     L_i(\omega) = B_i\left(\frac{\omega_{c,i}^2}{\omega_{c,i}^2+(\omega-d_i)^2}+\frac{\omega_{c,i}^2}{\omega_{c,i}^2+(\omega+d_i)^2}\right),\label{eq:SymLorentzian}
\end{equation}
depends on three parameters: $\{B_i, d_i,\omega_{c,i}\}$. Hence, $\boldsymbol{\theta}$ consists of $3N_{\rm basis}$ parameters. A Lorentzian basis is also physically meaningful, as environmental correlations are expected to decay exponentially in time~\cite{forster2018hydrodynamic, Lindsey1980}. 

Our variational algorithm for a set of $N_{\rm coh}$ coherence measurements, $\{ C_j(t)\}$ for $j \in [1, N_{\rm coh}]$, arising from the application of different pulse sequences, consists of:
\begin{enumerate} 
    \item Initializing $\boldsymbol{\theta}$ stochastically and constructing $S^{\rm trial}(\omega)$ from the sum of $N_{\rm basis}$ Lorentzians.
    
    \item Evaluating $\{ C_j^{\rm trial}(t)\}$ from $S^{\rm trial}(\omega)$ and $\{F_j(\omega t)\}$, with $j$ denoting all distinct pulse sequences.
    
    \item Calculating the loss, $\mathcal{L}(\boldsymbol{\theta})$, that quantifies the deviation of $\{ C_j^{\rm trial}(\boldsymbol{\theta}; t)\}$ from the measured $\{ C_j(t)\}$.
    
    \item Variationally optimizing $\boldsymbol{\theta}$ until reaching an error threshold, $\xi$, subject to the constraint $\boldsymbol{\theta} \geq 0$. 
\end{enumerate}
Upon converging, this protocol yields an optimal $\boldsymbol{\theta}^*$ that defines $S(\omega; \boldsymbol{\theta}^*)$ and minimizes the error in the resulting coherence, $C(t; \boldsymbol{\theta}^*)$, with the measured $C(t)$ (see SI~Sec~II for additional details on the algorithm and convergence criteria). Our VQNS thus employs data from multiple pulse sequences \textit{simultaneously} and \textit{self-consistently}. The protocol is valid for any $N_{\rm coh}$ but can be expected to work best when one uses $N_{\rm coh} > 1$ coherence measurements with \textit{complementary} sensitivities (see discussion on sensitivity measure). 

\begin{figure}
\hspace{-.5cm}{\includegraphics[width=0.5\columnwidth]{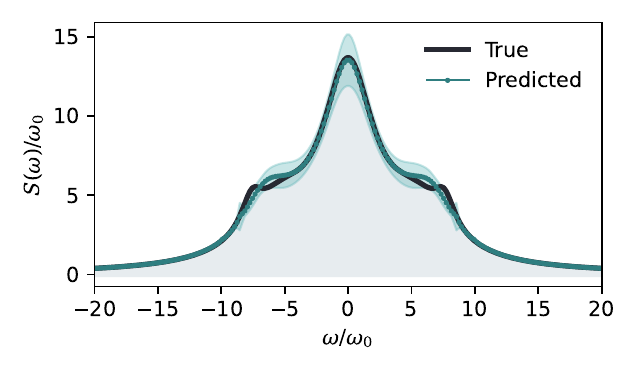}}
\vspace{-15pt}
\caption{VQNS reconstruction of a sample power spectrum from 7 input coherence measurements using 0, 1, 2, 3, 8, 16, and 32-pulse CPMG sequences, a convergence threshold of $\xi = 1\times 10^{-5}$, and confidence intervals calculated from $N_{\rm runs}=20$ runs. The solid teal line is the pointwise mean of the reconstructed spectra and the light band is their pointwise standard deviation (confidence intervals). The confidence intervals indicate the frequency regions over which the reconstructed spectrum can be modified, and by how much, for the resulting coherence curves to still agree with the input coherences within the same convergence threshold.
} 
\label{fig-basic}
\vspace{-15pt}
\end{figure}

An advantage of our approach is that the stochastic initialization of $\boldsymbol{\theta}$ yields slightly different optimized noise spectra for each run of the algorithm, allowing us to quantify confidence intervals for our noise spectrum predictions. In all our results, we provide the average noise spectrum (teal dots) obtained from $N_\mathrm{runs}$ independent runs of our algorithm subject to the same convergence threshold, $\xi$. In all tests, $N_{\rm runs} = 20$ converges the means. The confidence intervals (translucent teal) track the pointwise standard deviation of the $N_{\rm runs}$ predictions. 

We begin our analysis with theoretical spectra of increasing complexity. In Fig.~\ref{fig-basic}, we consider VQNS's ability to extract a structured noise spectrum consisting of a sum of three Lorentzian peaks from coherence curves obtained by applying 7 distinct pulse sequences. We employ a small basis of $N_{\rm basis}=3$ Lorentzian peaks in the optimization process. The confidence intervals in Fig.~\ref{fig-basic} correctly reflect the uncertainty in the shoulder feature at $\omega \sim 5$, where the mean predictions deviate slightly from the true spectrum and the fact that the central peak height can vary without significantly affecting overall agreement with input coherence curves. 

The deviation of our VQNS predictions from the true spectrum highlights the lossy transformation connecting power spectra to coherence curves. Like in Laplace transforms, the exponentially damped form of Eq.~\ref{eq:FFformalism} means that a \textit{family} of similar noise spectra can reproduce the same coherence curves within a small error threshold. Nonetheless, VQNS accurately captures the positions, widths, and heights of the constituent Lorentzians. Because our algorithm is trivially parallelizable and efficient (a single run takes $\lesssim 5$ minutes on a commonly accessible laptop), performing statistical analysis on the fly is easy and offers the benefit of quantifying confidence intervals---a feature that becomes essential when analyzing noisy coherence data.

\begin{figure*}
\hspace{-.5cm}{\includegraphics[width=\textwidth]{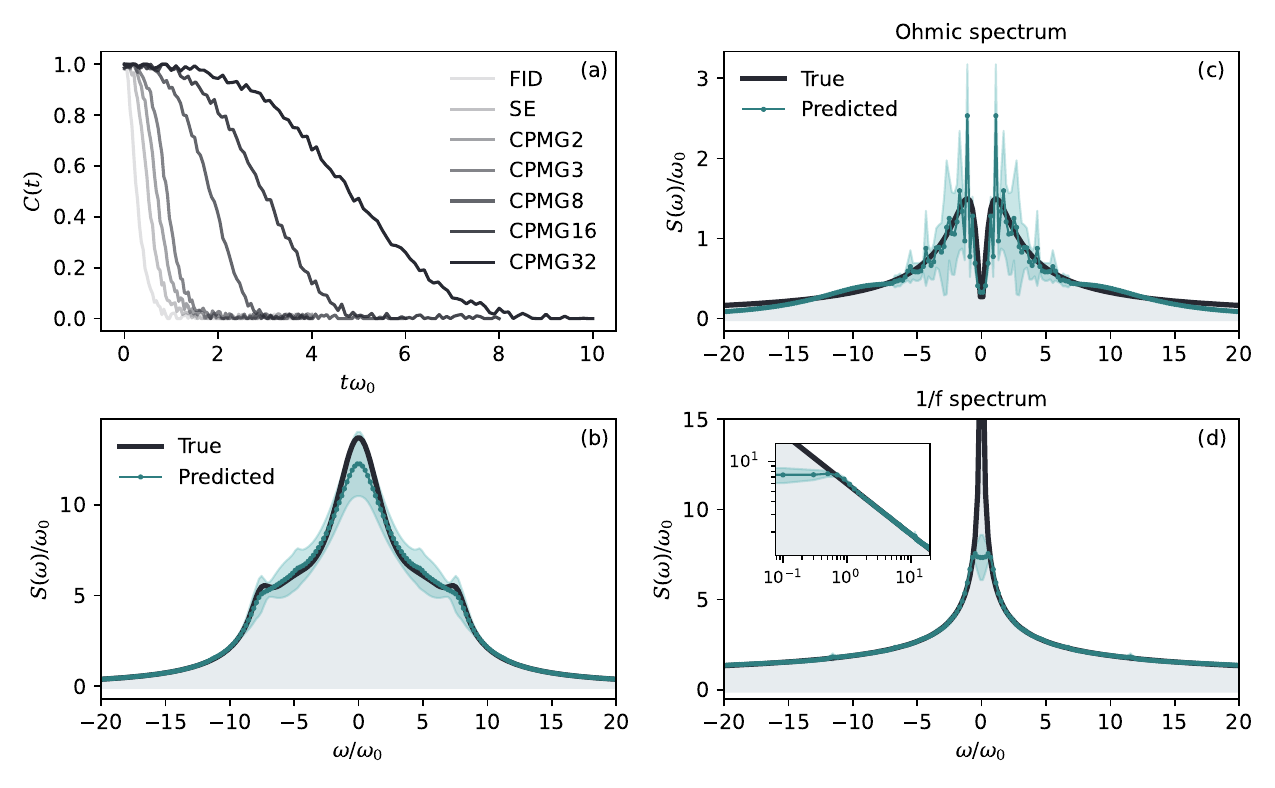}}
\vspace{-15pt}
\caption{(a) Numerically generated coherence dynamics with simulated measurement uncertainty of 2.0\% added on, at different strengths, and (b) the corresponding VQNS reconstruction result on these coherence inputs. We used a sample size of $N_{\rm runs}=20$ to quantify the confidence intervals. Right column: non-Lorentzian noise spectrum reconstruction of an (c) Ohmic spectrum, and a (d) $1/f$ spectrum. Panel (d) inset: same figure shown on a log-log scale. For both spectra, we used $\xi = 3\times 10^{-5}$, $N_{\rm basis}=20$, and $N_{\rm runs}=20$.} 
\label{fig-theoretical}
\vspace{-15pt}
\end{figure*}

Figure~\ref{fig-theoretical} tests VQNS's ability to extract noise spectra from measurements with statistical noise and poor temporal resolution, and notoriously challenging functional forms, like Ohmic spectra with a slow linear decay and a divergent 1/$f$ spectrum. In Fig.~\ref{fig-theoretical}~(a), we begin with coherence inputs with additive errors sampled from a uniform distribution spanning $[ -\epsilon, \epsilon)$, where $\epsilon=0.02$, mimicking uncorrelated experimental errors like background and shot noises~\cite{Hong2013}. Because this error in the coherences places an upper limit to the achievable loss, we adopt $\xi = 1.17\times 10^{-4}$, and use $N_{\rm basis}=10$ Lorentzian basis functions to obtain sufficient functional flexibility and shorten convergence time. Remarkably, the average prediction semiquantitatively agrees with the noise-free case in Fig.~\ref{fig-basic}, retaining the salient features of the true spectrum. The true spectrum also largely lies within the confidence intervals at all levels of measurement error. Thus, our VQNS offers a simple yet powerful means to extract noise spectra from realistic coherence curves, even when highly averaged data is not immediately available.

\begin{figure*}
\hspace{-.5cm}{\includegraphics[width=\textwidth]{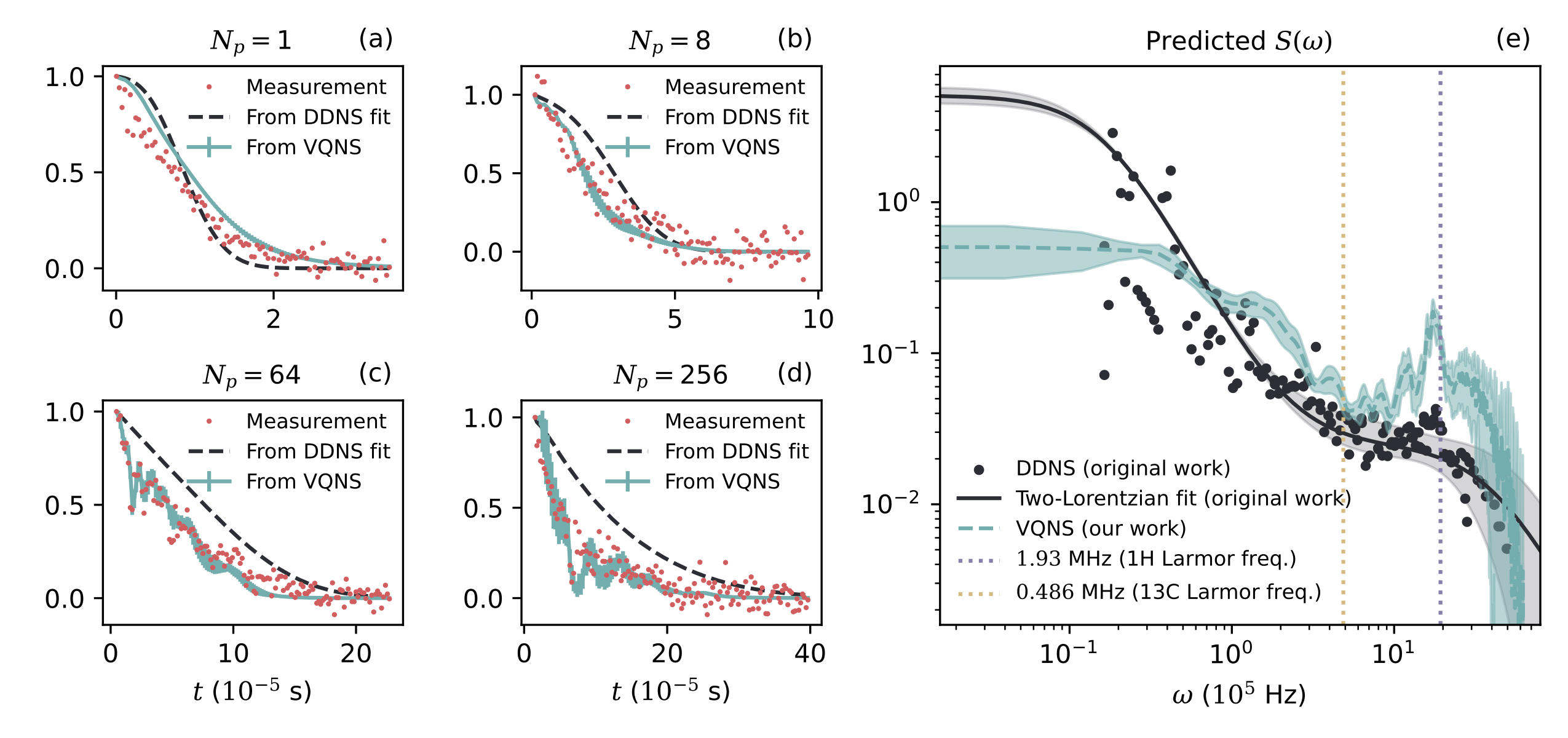}}
\vspace{-15pt}
\caption{(a-d) Comparison of experimentally measured coherence curves with those predicted using power spectra reconstructed using DDNS and VQNS for experiments with 1, 8, 64, and 256 CPMG pulses. We adopt $N_{\rm runs} = 30$, $\xi=1\times 10^{-2}$, $N_{\rm basis}=40$ Lorentzian basis functions, and set $\omega_0 = 10^{5}/(2\pi)$~Hz. . (e) Comparison of noise spectra reconstructed using DDNS taken from Ref.~\cite{Romach2015} (dark blue) and VQNS (teal). For the VQNS noise spectrum reconstruction, we show confidence intervals as a translucent teal band. Dotted vertical lines indicate the Larmor frequencies of hydrogen (purple) and carbon-13 (yellow) at an applied magnetic field of 454~G. The presence of a peak at the hydrogen Larmor frequency is expected due to the presence of protons from the immersion oil at the surface of the sample, in proximity to the NV center, and is also reported in the original study~\cite{Romach2015}. The lack of a feature appearing at the carbon-13 Larmor frequency is consistent with and likely due to the fact that the isotopically purified sample used contains $<0.001$\% carbon-13 impurities.} 
\label{fig-exptNV}
\vspace{-10pt}
\end{figure*}   

We now turn to the challenging Ohmic and $1/f$ spectral forms (see SI~Sec.~III). Employing seven CPMG-type pulse sequences with $N_{\rm p} = \{0, 1, 2, 3, 4, 5, 6\}$ with VQNS and a modest $N_{\rm basis} = 20$ basis for the Ohmic spectrum in Fig.~\ref{fig-theoretical}~(c) achieves good agreement within confidence intervals, including the slowly decaying high-frequency tails and suppressed contribution near $\omega \rightarrow 0$. For the $1/f$ spectrum in Fig.~\ref{fig-theoretical}~(d), we employ seven CPMG-type pulse sequences with $N_{\rm p} = \{1, 2, 3, 4, 8, 16,32\}$, omitting the 0-pulse measurement since its coherence (Eq.~\ref{eq:FFformalism}) diverges due to the unphysical $1/f$ behavior at $\omega=0$. Nevertheless, $1/f$-type power spectra are often invoked to describe the frequency scaling over \textit{finite regions in frequency space}, which our VQNS accurately captures~\cite{Paladino2014}. Interestingly, VQNS assigns a finite plateau near the unphysical $\omega \rightarrow 0$ limit (Fig.~\ref{fig-theoretical}~(d), inset), while yielding exquisite accuracy over the $\omega > \omega_0$ regime. These forms also reveal that VQNS is efficient and convergent, even in challenging cases (SI~Sec.~IV).

Having thus established the theoretical reliability of VQNS, we assess its ability to extract \textit{feature-rich} noise spectra directly from experiment. We employ all coherences ($N_{\rm p} = \{ 1, 2, 4, 8, 16, 32, 64, 96, 128,256
\}$) measured previously on shallow NV centers submerged in oil and used within DDNS to reconstruct the noise spectrum in Fig.~2~(e) of Ref.~\cite{Romach2015}. Figure~\ref{fig-exptNV} compares the previous DDNS spectral reconstruction \cite{Romach2015} to our VQNS results obtained when using all 10 available sequences (e), and coherence measurements to curves generated from the DDNS and VQNS noise spectra (a-d). To assess the accuracy of VQNS relative to DDNS, we examine which power spectrum more faithfully reproduces the experimentally measured coherence curves and resolves features corresponding to spin species expected to lie near the diamond surface (see also SI~Sec.~V for additional strategies to confirm the veracity of VQNS results). 
 
VQNS finds the low-frequency spectral region to be about ten times smaller than the double-Lorentzian reconstructed with DDNS in Ref.~\cite{Romach2015}, indicating the spectrum approaches zero frequency less steeply. A more accurate assessment of the zero-frequency spectral weight would require Ramsey measurements, which are sensitive to this limit but were not available in the original experiment \cite{Romach2015} (see our discussion of the sensitivity measure, below). Furthermore, the DDNS reconstruction exhibits a sharp cutoff around $\omega\sim0.1$~MHz. Approaching this cutoff from the right, the spread of the DDNS predictions span about two orders of magnitude, indicating large uncertainties in the DDNS-reconstructed spectrum as $\omega\rightarrow 0$. In contrast, the VQNS confidence intervals vary only by a factor of 1.5 to 2 for $\omega \leq 20$~MHz. These narrow confidence intervals of the VQNS results suggest a high level of accuracy, even in the low-frequency limit. Indeed, the good agreement between VQNS coherence curves and experiment---which is markedly better than that of the DDNS curves---across all CPMG pulse sequences shown in Fig.~\ref{fig-exptNV}~(a-d) confirms the accuracy of the VQNS reconstruction. This is consistent with the fact that the DDNS spectrum best reproduces the SE decay in Fig.~\ref{fig-exptNV}~(a) but progressively worsens in capturing the coherences for all higher pulse sequences in Fig.~\ref{fig-exptNV}~(b-d). 

The wealth of structure in our VQNS spectrum compared to the DDNS analogue is particularly striking. Specifically, the VQNS spectrum contains a prominent peak at $\approx 1.93$~MHz, corresponding to the hydrogen Larmor frequency at the experimentally applied magnetic field (454G)---a feature also reported in~\cite{Romach2015}, but which was only confirmed under a separate XY8 pulse sequence. This signal corresponds to hydrogen atoms from the immersion oil near the diamond surface. Further, the lack of any significant feature at the carbon-13 Larmor frequency ($\approx 0.5$~MHz) is consistent with the use of an isotopically purified diamond containing $<0.001\%$ carbon-13 impurities. Thus, our VQNS's ability to resolve the hydrogen feature from the CPMG measurements alone---a feat DDNS struggled to achieve---highlights its ability to extract greater physical insight from the same experiment.

\begin{figure*}
\hspace{-.5cm}{\includegraphics[width=\textwidth]{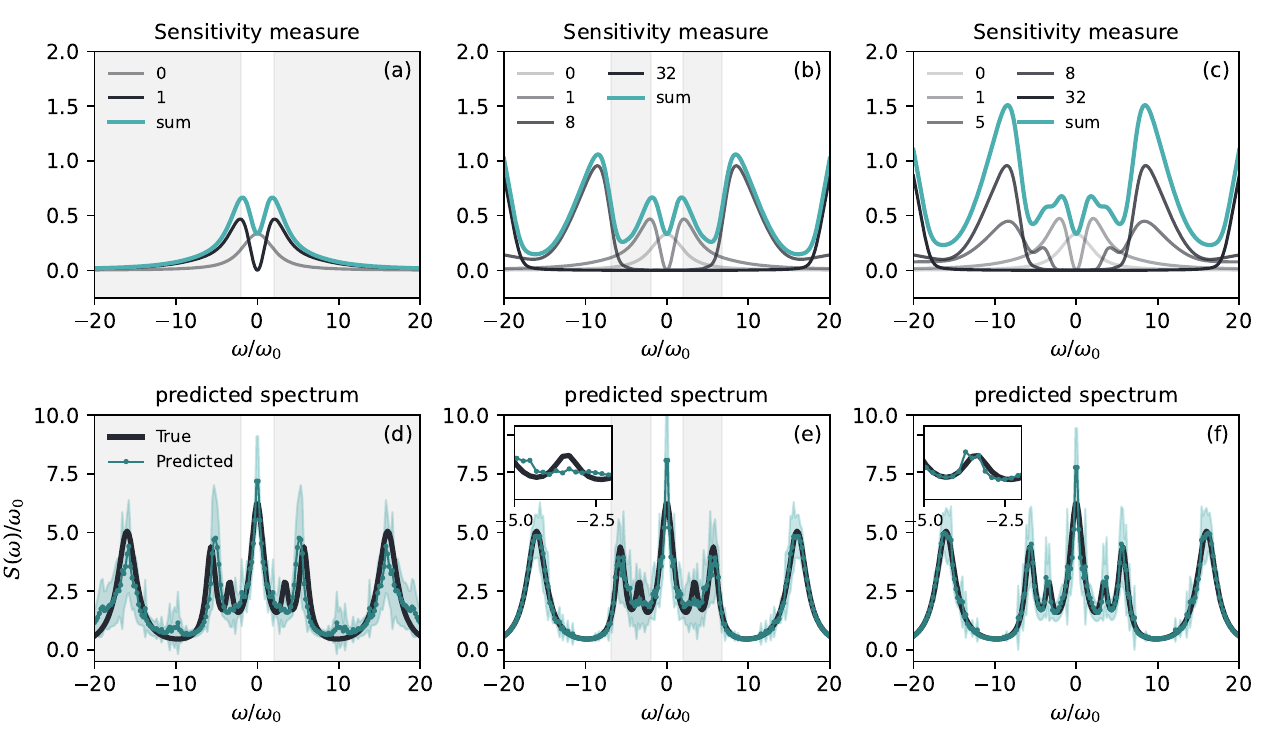}}
\caption{(a-c) Sensitivity measures $G_x(\omega)$ of pulse sequences used in the measurement set and their sum. (d-f) VQNS noise spectra reconstructed using the corresponding sets of measurements indicated in the above panels. For the spectrum reconstructions, we used $N_{\rm basis} = 20$, $N_{\rm runs}=20$, and $\xi = 1\times10^{-5}$.   The frequency regions of low sensitivity, i.e., low amplitude in the summed sensitivity measures, correspond to regions in the reconstructed spectra where the predictions vary significantly (shaded in gray), as indicated by the broad confidence intervals. Insets of (e) and (f) show a close-up view of the mean prediction and true spectrum values at the peak located at $\omega\approx -3.4\omega_0$, where the addition of the CPMG5 sequence in (f) improves the ability of the method to predict the presence of this peak compared to (e). 
{  For the coherence measurements used in the VQNS predictions and the evaluation of the sensitivity measures, 101 equidistantly spaced time points between initial times of 0 for all pulse sequences and a final time of $t_f \omega_0 \in \{3, 4, 5, 5, 6\}$ for pulse numbers $N_p \in \{0, 1, 5, 8, 32\}$ were used for all panels.}
}
\label{fig-FFanalysis}
\vspace{-10pt}
\end{figure*}

Beyond greater accuracy and efficiency, our VQNS can also suggest which experiment can best improve the precision of noise spectrum extractions. Specifically, Eq.~\ref{eq:FFformalism} implies that spectral reconstruction becomes more uncertain when the filter function of a particular pulse sequence becomes small. Hence, no single pulse sequence fully reveals the spectrum across all frequencies. One may, therefore, suppose that there exist \textit{complementary} sequences that, when taken together, uncover the noise spectrum across a wide frequency range. To find this complementarity, we propose a measure for a pulse sequence's sensitivity, $G_x(\omega)$, where $x$ denotes the pulse sequence, e.g., FID, SE, and $N_p$-pulse CPMG sequences. 

Physically, the \textit{sensitivity} of a pulse sequence depends on the time average of its instantaneous sensitivity, $F_x(\omega t)$, weighted by the signal, $C_x(t) = \exp[-\chi_x(t)]$,
\begin{equation}\label{eq:sensitivity}
    G_x(\omega) = \int_0^{\infty}dt\ F_x(\omega t)e^{-\chi_x(t)}.
\end{equation} 
However, since $\chi_x(t) = (4\pi)^{-1}\int_{-\infty}^{\infty} d\omega\ S(\omega)F_x(\omega t)$, this metric depends on the power spectrum one wishes to sense, $S(\omega)$. Hence, one must start with a preliminary set of measurements to extract a trial spectrum that can be used as input to $G_x(\omega)$ to predict the frequency sensitivity of other pulse sequences. We recommend starting with the simplest accessible measurements, i.e., FID and SE sequences, which probe the lowest frequency regions to obtain a preliminary $S(\omega)$. Using this $S(\omega)$, one can employ Eq.~\ref{eq:sensitivity} to determine the next set of pulse sequences that can optimally refine the confidence intervals associated with the VQNS reconstruction---a process that can be iterated to tighten the confidence intervals on the power spectrum reconstruction.

We test this refinement protocol on a structured test spectrum in Figure~\ref{fig-FFanalysis}. Fig.~\ref{fig-FFanalysis}~(a) illustrates the collective sensitivity of initially chosen FID and SE pulses, $G_\mathrm{FID}(\omega) + G_\mathrm{SE}(\omega)$, while Fig.~\ref{fig-FFanalysis}~(d) shows the extracted spectrum. The broad confidence intervals in the gray-shaded regions of the power spectrum in (d) overlap the region of low sensitivity in (a), confirming the validity of our sensitivity measure. This alignment suggests that pulse sequences with sensitivity in $\omega \in [3,20] \omega_0$ would increase the precision of the VQNS extraction. To test this idea, we include 8- and 32-pulse CPMG sequences which add sensitivity in the $\omega \in [8,18] \omega_0$ (c) and yield tighter confidence intervals in the same region of the reconstructed spectrum in (e). To further refine the spectral reconstruction in $\omega \in [2.5, 7] \omega_0$, which shows poor reconstruction accuracy in (e)-inset, we include the 5- and 32-pulse CPMG sequences in (c). This iterative refinement significantly improves the spectrum in (f), with tight confidence intervals throughout the entire frequency range. Thus, our protocol accurately identifies which experiment can best refine the power spectrum extraction. 

Our analysis also reveals a subtle danger: \textit{any} finite set of coherence measurements has regions with limited sensitivity. Since filter functions suppress the effect of spectral weight in these ``insensitive'' regions on the coherence curves, any spectral reconstruction technique could spuriously assign spectral weight to these regions without affecting the accuracy metric. We avoid such unphysical behavior via mild regularization (SI~Sec.~V). 

The feasibility of any noise spectroscopy, especially of single quantum sensors and information processors, lies in its ease of use, broad compatibility, and efficiency. To this end, VQNS---unlike DDNS and FTNS---accepts measurements under \textit{any} number of CPMG (and non-CPMG) pulse sequences, accessing a wider set of measurements under ``intermediate number'' DD pulses that can reliably be used to inform the predicted noise spectrum (SI~Sec.~VI). Our choices of variational basis and optimization algorithms---which permit fast evaluation of coherences and convergence in the variational manifold, respectively---ensure the efficiency of VQNS (SI~Sec.~IV details our cost scaling analysis). Indeed, the computational cost for this method is small, taking $\sim5-10$ minutes on a commercial laptop with 16GB RAM and an Apple M1 chip for a single run. 
Finally, the algorithm is trivially parallelizable, allowing one to quantify confidence intervals simultaneously across cores.

We have thus introduced a theoretical tool, VQNS, to post-process coherence measurements subject to arbitrary pulse sequences---commonly performed in quantum optics, chemistry, and atomic and molecular physics experiments around the world---and characterize the decoherence mechanisms of quantum systems. We established a variational principle that enables us to construct a fast, data-efficient, and statistical noise-robust tool to extract a power spectrum that \textit{simultaneously} and \textit{optimally} reproduces experimental measurements. Leveraging our VQNS, we proposed a refinement protocol to identify which measurements best refine spectral extractions, enabling an efficient theory-experiment loop. 

The success of VQNS in revealing the unexpected low-frequency noise structure and correctly resolving the $^1$H signal---which DDNS misses---and the absence of the $^{13}$C peak in the power spectrum in experiments on a NV sensor in isotopically purified diamond immersed in oil~\cite{Romach2015} suggests that a wealth of physical insight stands to be gained even from \textit{future} and \textit{currently available measurements} that are yet to be exploited. The accuracy, simplicity, and generality of our VQNS set the stage to achieve error-resilient and scalable precision quantum sensing across many quantum platforms. As the power spectrum is the central objective of quantum sensing protocols that aim to obtain a frequency-resolved characterization of the sensor-environmental coupling, we anticipate that VQNS will become a central tool in enabling precision quantum sensing. 

\begin{acknowledgement}

The authors acknowledge funding from the National Science Foundation (Grant No.~2326837). We thank Nir Bar-Gill for providing experimental data and helpful discussions, Bertram Cham for assistance with PyTorch, Shuo Sun and Joseph Zadrozny for insightful discussions, and Mikhail Mamaev and Zachary Wiethorn for comments on the manuscript.

\end{acknowledgement}

\begin{suppinfo}

Analytical expressions for response functions, details on the VQNS algorithm, implementation, convergence properties, general strategies for confirming the veracity of VQNS noise spectrum extractions, and comparisons to alternative methods. 

\end{suppinfo}

\bibliography{biblo-optNS} 

\end{document}


\title{Supplementary Information for \\``Fast, accurate, and error-resilient variational quantum noise spectroscopy''}
\author{Nanako Shitara}
\affiliation{Department of Chemistry, University of Colorado Boulder, Colorado 80309, USA}
\affiliation{Department of Physics, University of Colorado Boulder, Colorado 80309, USA}
\author{Andr\'es Montoya-Castillo}
\affiliation{Department of Chemistry, University of Colorado Boulder, Colorado 80309, USA}
\email{Andres.MontoyaCastillo@colorado.edu}
\date{\today}

\maketitle

\renewcommand{\theequation}{S\arabic{equation}}
\renewcommand{\thefigure}{S\arabic{figure}}
\renewcommand{\thepage}{S\arabic{page}}
\renewcommand{\bibnumfmt}[1]{$^{\mathrm{S#1}}$}
\renewcommand{\citenumfont}[1]{S#1}

\onecolumngrid

\vskip-10pt
\section{Attenuation functions for Lorentzian spectra subject to CPMG pulses}
\label{app:chiCPMG}

Here, we provide expressions for the attenuation function for a symmetrized Lorentzian function (as given in Eq.~1 in the main manuscript), restricted to CPMG pulse sequences ($N_p \geq 1$), as a function of $N_p$ and the parameters of the Lorentzian. $\chi_{\rm CPMG}^{\rm even}$ refers to the expression for the case where $N_p$ is even and $\chi_{\rm CPMG}^{\rm odd}$ where $N_p$ is odd. As a reminder, the attenuation function $\chi(t)$ is evaluated by the expression
\begin{equation}
    \chi(t) = \frac{1}{4\pi}\int_{-\infty}^\infty d\omega \, S(\omega) F(\omega t)
\end{equation}
where the filter functions $F(\omega t)$ for FID and CPMG sequences are given by
\begin{eqnarray}
    F_\mathrm{FID}(\omega t) &=& \frac{4}{\omega^2} \sin^2(\omega t/2), \nonumber \\
    F_\mathrm{CPMG, even}(\omega t) &=& \frac{16}{\omega^2} \frac{\sin^4(\omega t/4 N_p)}{\cos^2(\omega t/2 N_p)} \sin^2(\omega t/2), \nonumber \\
    F_\mathrm{CPMG, odd}(\omega t) &=& \frac{16}{\omega^2} \frac{\sin^4(\omega t/4 N_p)}{\cos^2(\omega t/2 N_p)} \cos^2(\omega t/2). \nonumber
\end{eqnarray}

The attenuation functions for CPMG sequences take the following expressions~\cite{Szankowski2018}:
\begin{eqnarray}
    \chi_\mathrm{CPMG}^{\rm even}(t, N_p) &=& \frac{B t \omega _c^2}{\omega _c^2+d^2} + \sum_{z \in \{ z_{-}, z_{+}\} }\frac{i B \omega _c \left(N_p \tan \left(\frac{t z}{2 N_p}\right)+4
   e^{-\frac{1}{2} i t z} \sin \left(\frac{1}{2} t z\right) \sin^4\left(\frac{t
   z}{4 N_p}\right) \sec^2\left(\frac{t z}{2 N_p}\right)\right)}{z^2} , \label{eq:chiEven}
\end{eqnarray}
and
\begin{eqnarray}
    \chi^{\rm odd}_\mathrm{CPMG}(t, N_p) &=& \frac{B t \omega _c^2}{\omega_c^2+d^2} + \sum_{z \in \{ z_{-}, z_{+}\} } \frac{B \omega_c \left( \left( 1-e^{\frac{itz}{2 N_p}}\right)^4 \left(1 + e^{-itz} \right) - 2 N_p \left(1 - e^{\frac{2 i t z}{N_p}}\right) \right)}{2 \left( 1 + e^{\frac{itz}{N_p}} \right)^2 z^2} ,\label{eq:chiOdd}
\end{eqnarray}
where $z_{+} \equiv d+i\omega_c$ and $z_{-} \equiv d-i\omega_c$.

The portability and efficiency of our VQNS are ensured by our choice of Lorentzians as our variational basis and adoption of state-of-the-art optimization algorithms that have now become the norm in machine learning approaches. We opt for a Lorentzian basis because of the need to efficiently evaluate the loss function in our variational procedure, which requires one to repeatedly calculate the coherence decays that arise from the application of a set of pulse sequences to a trial power spectrum (see Eq.~1). The availability of analytical expressions for the coherence response due to an arbitrary Lorentzian power spectrum subjected to any applied pulse sequence~\cite{Szankowski2018} obviates the need for numerical quadrature, rendering the evaluation of our loss function highly efficient. We note, however, that one may also consider different basis functions, such as Gaussians~\cite{Szankowski2018} or purely positive polynomials~\cite{Weierstrass1885} (albeit at the cost of lower efficiency for functions which do not have analytical expressions available). 

Beyond the technical advantage of enabling analytical integration of the coherence response, the Lorentzian basis also offers a physically inspired choice. Specifically, in dissipative systems, the environmental autocorrelation function whose Fourier transform defines the noise spectrum $S(\omega)$ can be expected to decay according to a combination of exponential functions~\cite{forster2018hydrodynamic, Lindsey1980}. In turn, the Fourier transform of a sum of exponential functions is given by a sum of Lorentzian functions. Thus, expressing the noise spectrum as a sum of Lorentzian functions is consistent with the expected behavior of the environmental autocorrelation function in the dissipative systems of interest to quantum noise spectroscopy.

\section{Variational algorithm, optimizer settings, and convergence}
\label{app:optim-details}

Here, we provide additional details on the variational algorithm, the optimizers, and hyperparameters we used to generate all figures in this manuscript. We also outline general strategies to satisfy reproducibility criteria and select appropriate convergence thresholds. We begin by augmenting our discussion of the algorithmic steps:
\begin{enumerate} 
    \item \textit{Initializing $\boldsymbol{\theta}$ stochastically and constructing $S^{\rm trial}(\omega)$ from the sum of $N_{\rm basis}$ Lorentzians.} We initialize $\boldsymbol{\theta}$ from uniform distributions: $B_i \equiv \tilde{B}_i/\omega_0 \in [0, 10]$; $\omega_{c,i} \equiv \tilde{\omega}_{c,i}/\omega_0 \in [0.1, 10]$; and $d_i \equiv \tilde{d_i}/\omega_0 \in \delta d[i, i+1]$, where $\delta d = d_{\rm max} /N_{\rm basis}$ and $d_{\rm max}/\omega_0 = 20$. We choose $\omega_0$ such that the filter functions span a frequency range of $\sim 20 \omega_0$. Imposing these conditions improves the efficiency and performance of the parameter optimization. 
    
    \item \textit{Evaluating $\{ C_j^{\rm trial}(t)\}$ from $S^{\rm trial}(\omega)$ and $\{F_j(\omega t)\}$, with $j$ denoting all distinct pulse sequences.}
    
    \item \textit{Calculating the loss, $\mathcal{L}(\boldsymbol{\theta})$, that quantifies the deviation of $\{ C_j^{\rm trial}(\boldsymbol{\theta}; t)\}$ from the measured $\{ C_j(t)\}$.} $\mathcal{L}(\boldsymbol{\theta})$ corresponding to the sum of the mean squared errors (MSE) of the true and trial coherence functions integrated over the measurement times and normalized by the total experimental times $T_j$. For discrete-time measurements, $\mathcal{L} = N_t^{-1}\sum_{j,k}  dt\ |C_j(t_k) - C_j^{\rm trial}(t_k)|^2$, where $t_k$ are the discrete times over which the measurement was performed and $N_t$ is the total number of temporal measurements. 
    
    \item \textit{Variationally optimizing $\boldsymbol{\theta}$ until reaching an error threshold, $\xi$, subject to the constraint $\boldsymbol{\theta} \geq 0$.} We employ a gradient-based optimizer and iterate to convergence, defined by an error threshold, $\xi$, subject to the constraint $\boldsymbol{\theta} \geq 0$. We use the Adam or AdamW optimizers, which converge efficiently through adaptive learning rates, underlie optimization protocols in machine learning~\cite{pytorch, tensorflow}, and are accessible via PyTorch~\cite{pytorch}. 
\end{enumerate}

We implement our variational algorithm using PyTorch~\cite{pytorch}, which provides built-in functions for efficient optimization and offers a convenient framework to set up our iterative procedure. The computational cost required for this method is small, generally taking $5-10$ minutes on a commercial laptop with 16GB RAM and an Apple M1 chip for a single run. Because the algorithm is trivially parallelizable, one can perform independent runs to estimate statistics and quantify confidence intervals simultaneously across cores. As an illustrative example, we note that the individual optimization that took the longest time ($\sim 30$ minutes) was for Fig.~3~(c).

For all figures, we employed our PyTorch implementation of an iterative optimization scheme using the Adam or AdamW optimization algorithm. The iterative minimization of the loss function repeats until the MSE loss value falls below the $\xi = 1\times 10^{-5}$ threshold. If the loss fails to converge to the required threshold within 10000 iterations, we terminate the procedure and restart it with new initial parameters.

\begin{figure*}
\hspace{-.5cm}{\includegraphics[width=\textwidth]{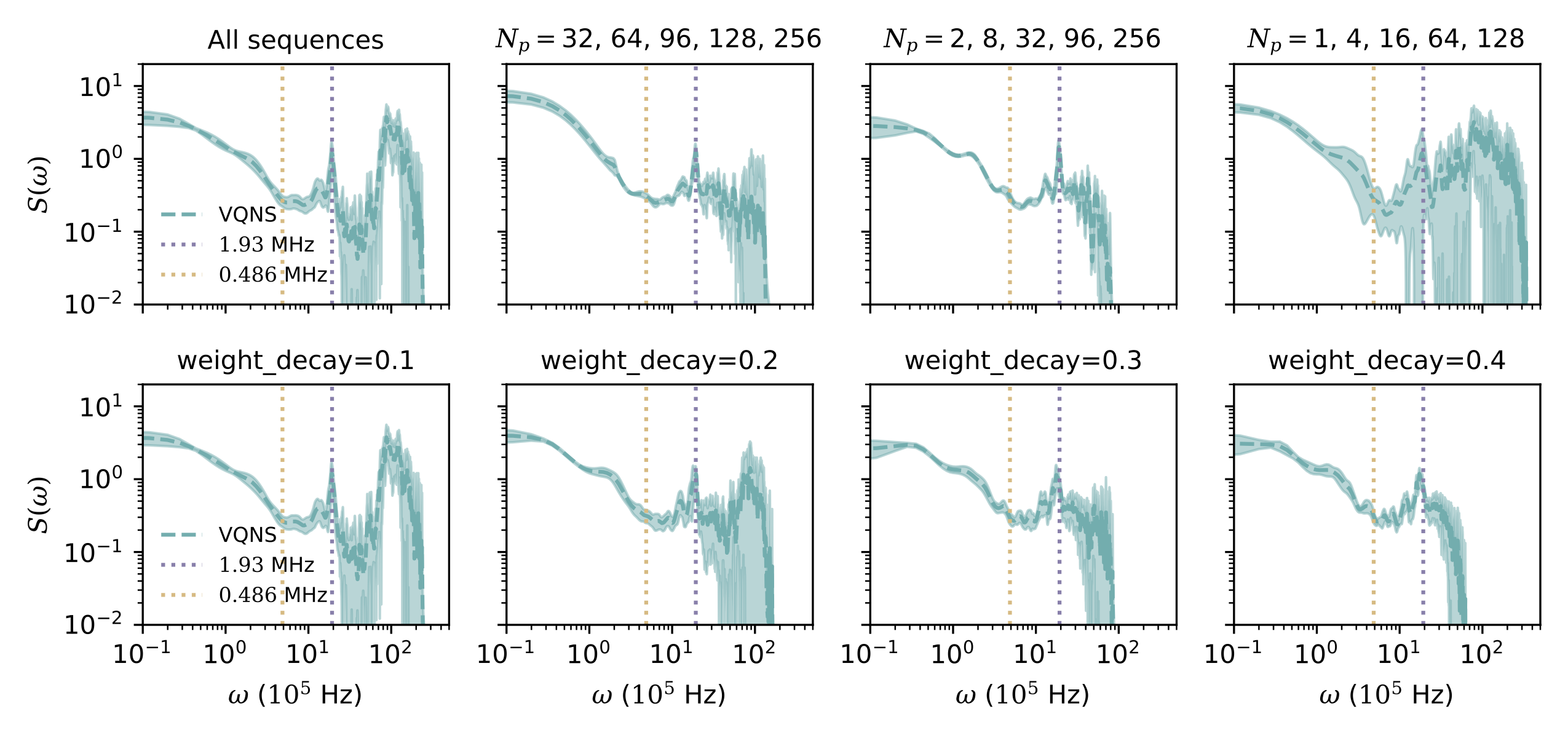}}
\vspace{-10pt}
\caption{VQNS predictions of the noise spectrum of a NV quantum sensor in diamond from experimental measurements reported in Ref.~\cite{Romach2015} and used in Fig.~4 in the main text, using different subsamples of the available measurements as input (top row), and varying the regularization strength in the optimizer (bottom row). Across both tests, the peak at the hydrogen Larmor frequency (indicated by the purple dashed vertical line) appears consistently, suggesting its veracity. Also across both tests, an additional structure at higher frequency regions ($>3\times 10^5$ Hz) and at the carbon-13 Larmor frequency (indicated by the yellow dashed line) fail to manifest consistently, indicating their possible spurious nature.} 
\label{fig-expt-supp}
\vspace{-10pt}
\end{figure*}


We opted for the \texttt{Adam} optimizer to generate  Figs.~2, 3~(a) and (b), and 5, with the learning rate hyperparameter \texttt{lr=0.01}. For Fig.~3~(c) and (d), we used \texttt{AdamW}, which decouples the weight decay in the regularization from the optimization steps, offering a more efficient implementation of regularization~\cite{Loshchilov2017}. We use the hyperparameters \texttt{lr=0.01}, \texttt{eps=1e-6}, \texttt{weight\_decay=0.01}, and \texttt{betas=(0.9, 0.9)}. For Fig.~4, we use the \texttt{AdamW} optimizer with hyperparameters \texttt{lr=0.02} and \texttt{weight\_decay=0.4}. For Figs.~\ref{fig-FTNScomp} and~\ref{fig-FTNScomp-addCPMG}, we used the \texttt{AdamW} optimizer with hyperparameters \texttt{lr=0.01} and \texttt{weight\_decay=0.1}.

To ensure the reproducibility of the reconstructed noise spectra, the optimization process should converge with increasing flexibility and the number of variational parameters (i.e., the size of the basis used to reconstruct the spectrum, $N_{\rm basis}$). Consequently, all our predictions are in the limit where the results converge with respect to $N_{\rm basis}$ subject to a chosen iterative error threshold, $\xi$. To achieve this, it is beneficial to perform reconstructions with increasing $N_{\rm basis}$, check that each choice of $N_{\rm basis}$ can satisfy the error threshold, $\xi$, and confirm that the reconstructed spectra from these runs fall within the target confidence intervals. 

To choose the convergence threshold, $\xi$, which allows one to enhance the signal-to-noise ratio, one should perform simple preliminary tests that account for the level of experimental uncertainty in the available measurements. Specifically, one should choose an initially conservative estimate (e.g., $\xi^{\rm trial} = 10^{-5}$) based on the level of uncertainty present in the input signals, and run the optimization a few times. If the loss function plateaus at a consistent but higher value, $\xi^{\rm new} > \xi^{\rm trial}$, throughout these tests, one should adopt $\xi^{\rm new}$ as the new convergence threshold for efficient performance. Conversely, if the optimizations easily reach $\xi^{\rm trial}$, one should lower this threshold to quantify the level of agreement the optimizer can effectively achieve. Generating confidence intervals based on these runs indicates whether the applied convergence threshold is sufficiently small to give reliable predictions.


\section{Non-Lorentzian test spectra parameters}
\label{app:nonLor-details}

The Ohmic spectrum in Fig.~3~(c) takes the form, 
\begin{equation}
    S_\mathrm{Ohmic}(\omega) = \eta \frac{\vert\omega\vert}{\omega^2 + \gamma^2}  \label{ohmicFunc}
\end{equation}
with $\gamma = 1.1 \omega_0$ setting the timescale of decay and $\eta = 3.3\omega_0$ setting the qubit-environmental coupling energy. 

The divergent $1/f$ spectrum in Fig.~3~(d) is given by
\begin{equation}
    S_{1/f}(\omega) = \frac{\zeta}{\vert \omega\vert } \label{1overfFunc}
\end{equation}
with $\zeta = 6\omega_0$ quantifying the qubit-environmental coupling. 


\section{Scaling of runtime and basis set size with convergence threshold}
\label{app:scaling}

In this section we discuss general trends of the dependence of the VQNS runtime with a specified convergence threshold, as well as of the dependence of the number of basis functions $N_\mathrm{basis}$ required for convergence with a specified convergence threshold. In both cases, we investigate two representative noise spectra, one with an ``easy'' structure and one with a ``complex'' structure, to provide a complete picture of how these trends may vary with the complexity of the underlying spectrum.


\begin{figure*}
\hspace{-.5cm}{\includegraphics[width=\textwidth]{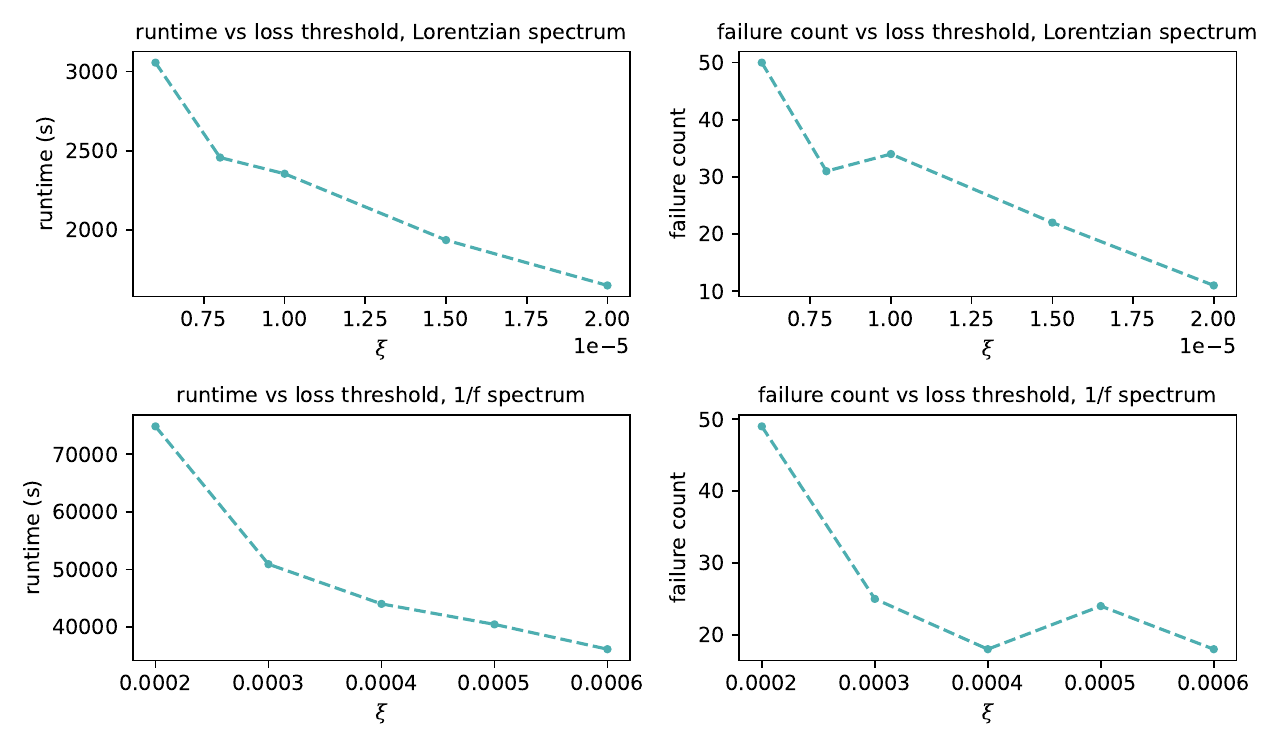}}
\vspace{-15pt}
\caption{Scaling of the total time to complete 100 successful optimizations of parameter values to the required threshold values for two representative spectra, the three-Lorentzian spectrum in Fig. 2 of the main text, and the $1/f$ spectrum in Fig. 3(b) of the main text. Because the maximum number of iterations (set to 1000 for the Lorentzian spectrum and 5000 for the $1/f$ spectrum) to take within which convergence must be reached for the run to be considered successful is fixed, the number of ``failed'' iterations are also plotted alongside each. It can be seen, as expected, that the runtime trend roughly follows that of the number of failures encountered within the sets of runs.} 
\label{fig-runtimeconv}
\vspace{-10pt}
\end{figure*}


We show in Fig.~\ref{fig-runtimeconv} the scaling of the total runtime of the VQNS method succeeding at 100 optimizations (including the runtime of any failed cycles, where the algorithm fails to converge to the required threshold within 1000 iterations for the three-Lorentzian spectrum, and within 5000 iterations for the $1/f$ spectrum) with the convergence threshold value, and also the scaling of the number of failed cycles with the convergence threshold. A large number of optimization runs were used to reflect a statistically accurate picture of the scaling of runtime with loss threshold. These scaling trends are shown for two representative noise spectra, the three-Lorentzian spectrum shown in Fig.~2 of the main text, and the $1/f$ spectrum shown in Fig.~3(d) of the main text. We performed all benchmarking on a multi-core commercial Mac mini (2020) with 16GB RAM and an Apple M1 chip. 

Several qualitative observations emerge. First, and unsurprisingly, the complexity of the underlying spectrum plays a significant role in the average runtime of each optimization run. For example, the average runtimes for the $1/f$ spectrum are an order of magnitude larger than that for the three-Lorentzian spectrum. Similarly, for threshold values, one also sees about an order of magnitude larger runtimes than those achieved for the three-Lorentzian spectrum. Second, as one may have anticipated, both spectra show that the runtimes decrease as the loss threshold is increased. What is more, the way in which they decrease follows closely the way in which the number of failed runs scale with loss threshold for both spectra. This makes intuitive sense since the failed runs are those that take the longest time to complete since they saturate the maximum number of iterations allowed per run. Thus, the practical convergence time increases with decreasing loss threshold in a way that depends on the underlying spectrum complexity, and through the algorithm being able to attain convergence less frequently within a fixed number of maximum iterations per run. 

We now turn to a more quantitative analysis of average runtimes for these two example spectra including and without including failures. First, the average time taken for a single failed run for the Lorentzian spectrum (with maximum 1000 iterations) was about 26.15 seconds, and for the $1/f$ spectrum (with maximum 5000 iterations) was about 571.09 seconds. Thus the average runtime of a single run, including failed runs, was $3055.38/100 = 30.55$ seconds for the Lorentzian spectrum, and $74816.31/100 = 748.16$ seconds for the $1/f$ spectrum, for the lowest threshold values considered for both cases. The true average runtime, representative of the average time a successful run would take to complete, would be closer to an average of the runtime excluding the estimated total time taken on failed runs, which would be $(3055.38 - 50 \times 26.15)/100 = 17.48$ seconds for the Lorentzian spectrum, and $(74816.31 - 50 \times 571.09)/100 = 462.62$ seconds for the $1/f$ spectrum, again for the lowest convergence threshold considered for both. Hence, even for challenging spectra, our VQNS offers an efficient approach to noise spectrum extraction.


\begin{figure*}
\hspace{-.5cm}{\includegraphics[width=\textwidth]{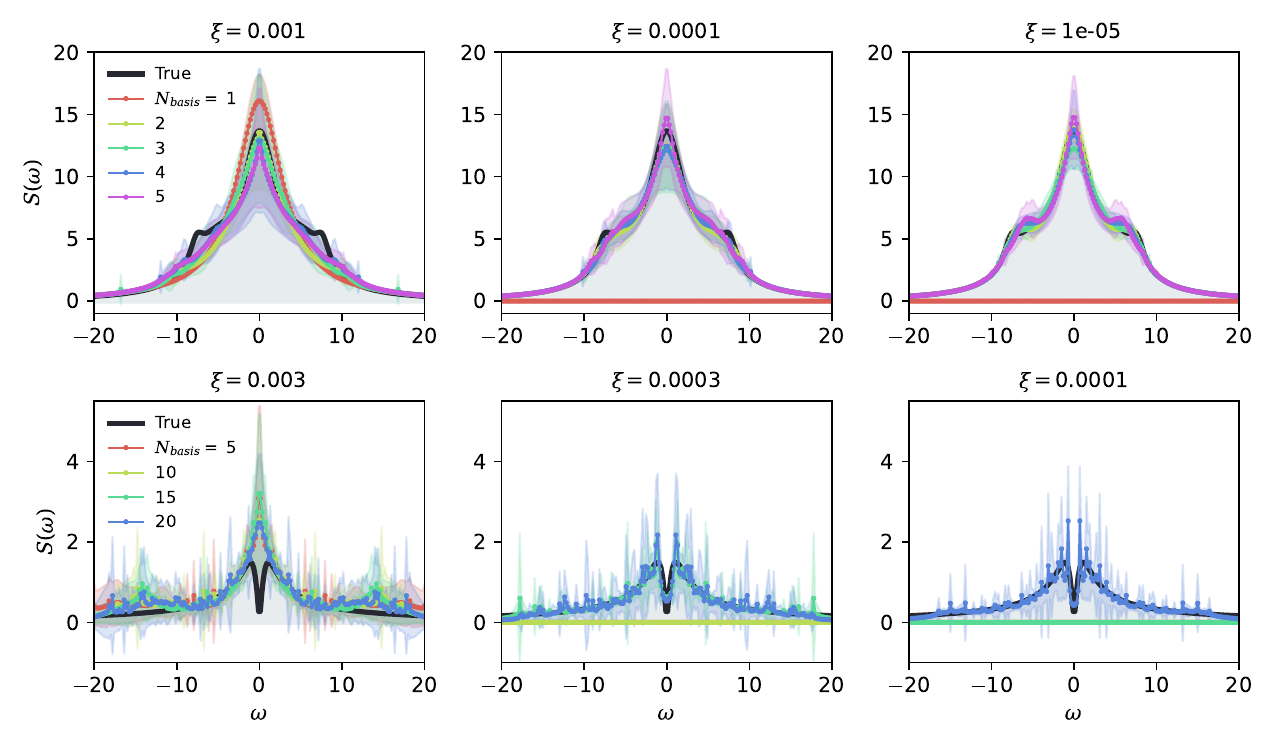}}
\caption{The results of applying VQNS on two representative spectra, the three-Lorentzian spectrum in Fig. 2 of the main text, and the Ohmic spectrum in Fig. 3(a) of the main text, with varying numbers of basis Lorentzian functions. The results where the spectrum is zero across all frequency indicates runs for which convergence was not attained, as defined by the criteria described in the main body of the response text. As can be seen, the most important factor in deciding the number of required basis functions to employ for efficient convergence is the overall complexity of the underlying spectrum, although setting a lower convergence threshold can lower this limit. The maximum number of iterations per run was set to 1000 for the Lorentzian spectrum and to 10000 for the Ohmic spectrum.} 
\label{fig-Nbasisconv}
\vspace{-10pt}
\end{figure*}


We now discuss the dependence of required number of Lorentzian basis functions for convergence subject to a convergence threshold (see Fig.~\ref{fig-Nbasisconv}). Unsurprisingly, convergence with basis set size again depends strongly on the complexity of the structure of the underlying noise spectrum. To test this dependence, we again take the three-Lorentzian spectrum from Fig.~2 as a representative ``simple'' example, and the Ohmic spectrum from Fig.~3(c) as a representative ``complex'' example. In principle, as long as there is sufficient probability for the optimization algorithm to reach the specified convergence threshold within the specified maximum number of iterations, one can run the algorithm indefinitely until the required number of successful runs is achieved. To simplify the analysis, we have arbitrarily defined the criteria for successful convergence as being able to complete the $N_\mathrm{runs}$ required runs of the optimization successfully before encountering $N_\mathrm{runs}/2$ failed runs. For the plots shown in Fig.~\ref{fig-Nbasisconv}, we have specified for both spectra $N_\mathrm{runs} = 20$, so that if a given set of runs reaches 10 failures before reaching 20 successes, the number of basis functions $N_\mathrm{basis}$ employed in this set of runs is considered insufficient for convergence. As Fig.~\ref{fig-Nbasisconv} shows, $N_\mathrm{basis}=2$ is the minimum number sufficient to converge the optimization algorithm for all loss threshold values considered for the simple three-Lorentzian spectrum. For the largest threshold threshold value considered, $\xi=0.001$, even $N_\mathrm{basis} = 1$ is sufficient for convergence. One should also consider convergence at the level of consistent results between different runs with different $N_\mathrm{basis}$. For $\xi=0.001$, although all values of $N_\mathrm{basis}$ ``converge'' within a fixed value of $N_\mathrm{basis}$, the average spectrum reconstructions only start agreeing well from about $N_\mathrm{basis}=4$ onward (although all averages lie more or less within confidence intervals of each other). In comparison, for the lower threshold results, all results that converged within fixed $N_\mathrm{runs}$ values show average predictions that agree well with each other, even starting from $N_\mathrm{basis} = 2$.

Moving to the Ohmic spectrum, we immediately see that the range of $N_\mathrm{basis}$ values to consider increase significantly. For the representative $\xi$ values considered, we see that as low as $N_\mathrm{basis}=5$ is sufficient to converge (within fixed $N_\mathrm{basis}$) for $\xi=0.003$, but at least $N_\mathrm{basis}=15$ is required for $\xi=0.0003$, and at least $N_\mathrm{basis}=20$ is required for $\xi=0.0001$. For these examples, all average predictions (for a fixed $\xi$ and varying $N_\mathrm{basis}$) are consistent with each other (this is of course not seen for the $\xi=0.0001$ case since only the $N_\mathrm{basis} = 20$ instance converged). These test cases suggest that the lower threshold predictions give average predictions that are more faithful to the true underlying spectrum. Hence, the number of basis functions required for convergence increases with decreasing loss thresholds. The particular values of the required number of basis functions depends on the complexity of the underlying spectrum, with a spectral structure deviating from simple sums of Lorentzian functions requiring more basis functions to reconstruct. Convergence with respect to consistent predictions across varying $N_\mathrm{basis}$, however, seems to be determined largely by the value of the loss threshold. That is, for a sufficiently low loss threshold, the predicted spectra are consistent with each other, even from the lowest number of $N_\mathrm{basis}$ required for individual convergence, providing a concept of a sufficient level of loss threshold value to achieve. 

In summary, we demonstrate that the runtime and the basis set size required for achieving converged parameters consistently within a fixed number of maximum iterations depend primarily on the complexity of the underlying noise power spectrum. Explicitly non-Lorentzian noise spectra, such as of the $1/f$ and Ohmic forms take both longer in runtime and require a larger number of basis functions to attain converged parameter values with a reliable probability per iteration ($\geq$50\% for the basis number test shown in Fig.~\ref{fig-Nbasisconv}). The specific runtime and basis function numbers required depend on the value of the loss threshold, with a lower threshold demanding both higher runtimes and basis function numbers. For any given set of runs, the total runtime may be impacted by the number of failed runs encountered. These failed runs have the longest runtimes among all runs, so it is best to determine which combination of maximum iteration number, loss threshold value, basis function number, and optimizer parameters leads to a reasonable success rate of the runs through preliminary testing. We also demonstrate that the predicted spectra can be tested for convergence with respect to the number of basis functions. In particular, we show that agreement in the predictions between different basis numbers indicate convergence within a selected loss threshold, and that agreement in the predictions between different loss thresholds indicates that a sufficiently low loss threshold has been chosen for the particular sets of measurements employed.

\section{Confirming VQNS analysis of experimental data}
\label{app:exptMisc}

Here, we outline strategies for verifying the predicted features obtained using the VQNS method when one is limited to an available set of measurements. An accessible approach would be to study the predictions made on different subsets of the available measurements to confirm which features are predicted robustly across different subsamplings. Features that do not consistently appear across different subsamplings indicate that these may not be reliable. Furthermore, when accessible measurements display high measurement uncertainties, we recommend adjusting the level of regularization implemented in the optimizer to identify and suppress features that may be a consequence of the measurement uncertainty rather than the true signal. 

\begin{figure*}[t]
\hspace{-.5cm}{\includegraphics[width=0.8\textwidth]{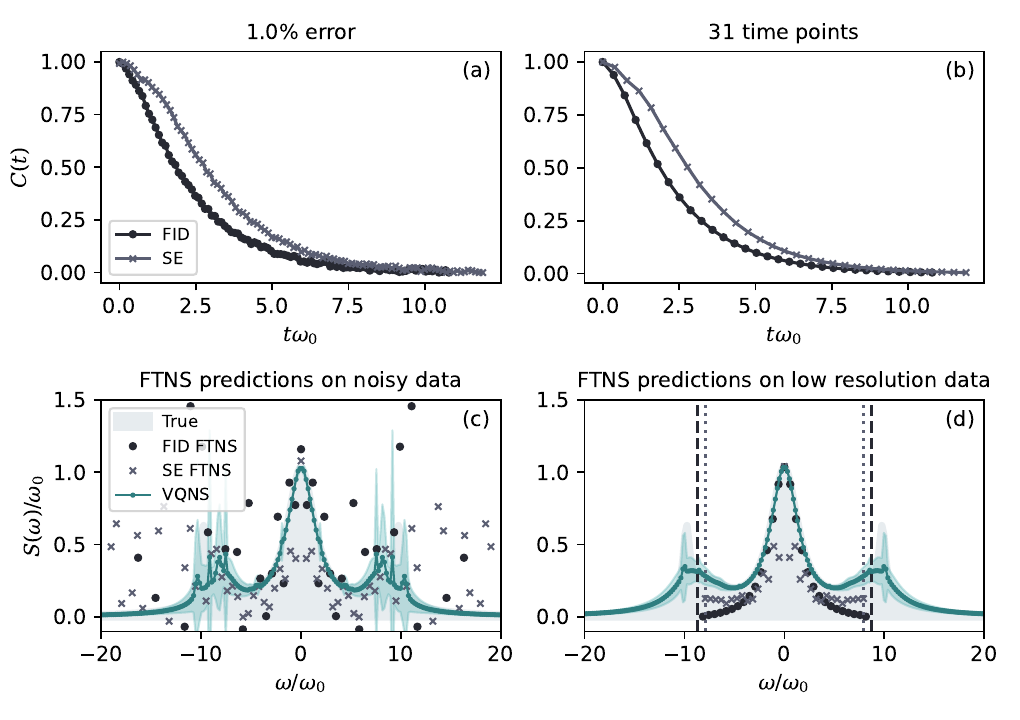}}
\vspace{-10pt}
\caption{(a, c) Comparison of FID FTNS, SE FTNS, and VQNS with FID and SE measurements with 1.0\% added simulated measurement noise. FTNS methods are sensitive to measurement uncertainties, whereas VQNS shows robustness against these. We employ a loss threshold of $\xi=3.5\times10^{-5}$, $N_{\rm basis}=3$, and $N_{\rm runs}=20$. (b, d) Comparison of FID FTNS, SE FTNS, and VQNS power spectrum reconstructions for coherence input with 31 total measurement points. For FTNS, the temporal resolution of measurements directly translates into limits of frequencies up to which the spectrum reconstruction can be performed reliably (dashed and dotted vertical black lines in (d)) but VQNS can overcome such limitations even for low measurement resolution. For these data, we use a loss threshold of $\xi=1\times10^{-6}$, $N_{\rm basis}=3$, and $N_{\rm runs}=20$.} 
\label{fig-FTNScomp}
\vspace{-10pt}
\end{figure*}   


In Figure~\ref{fig-expt-supp}, we show the results of applying both tests to the experimental data used in Fig.~4 of the main text. The first row of the figure shows spectrum predictions obtained from the application of VQNS with the \texttt{AdamW} optimizer with hyperparameters \texttt{lr=0.02} and \texttt{weight\_decay=0.1} to different subsets of the 10 available CPMG measurements. Interestingly, while all subsets consistently predict the presence of the prominent hydrogen Larmor frequency peak (purple vertical dashed line), the broad higher-frequency feature only appears in two of the four cases considered. This indicates that the structure predicted in the highest frequency region may not be as reliable. To further test this, in the second row of the same figure, we show the results of VQNS predictions now using \textit{all} CPMG measurements as input, but varying the \texttt{weight\_decay} hyperparameter, which controls the strength of the regularization applied in the optimization. As the regularization strength increases, the suspected spurious high-frequency structures are gradually suppressed, confirming that they may not be features of the true noise spectrum. Importantly, in both strategies, the signal at the hydrogen Larmor frequency persists, albeit at a slightly reduced frequency value as the regularization strength is increased. This red shift of the hydrogen signal is a consequence of applying a stronger regularization, whose effect is similar to the application of an effective smoothing procedure that biases observed frequency features towards a lower value. Furthermore, the inconsistent appearance of small features around the carbon-13 Larmor frequency (yellow dashed line) across these tests indicates the absence of sufficiently proximal carbon-13 nuclei from the target NV center, despite some panels showing small peaks that may, at first glance, appear as meaningful signals.

Based on these results, we have used the VQNS prediction using \texttt{weight\_decay=0.4} for the final predicted spectrum in Fig.~4 of the main text, which displays suppressed high-frequency features, as the most conservative reconstruction that reproduces the minimal spectral features required to optimize agreement in the resulting coherence.

\section{Comparison of VQNS with other methods}
\label{app:VQNS-discussion}

While other noise spectroscopy methods aim to exploit multiple coherence measurements simultaneously and self-consistently~\cite{Sun2022, Soetbeer2021}, our VQNS offers greater generality and systematic improvability. For example, Ref.~\cite{Sun2022} {does not employ an explicit optimization algorithm to achieve the self-consistent prediction. It also does not make use of the analytical expression of the coherence response obtained from a sum of Lorentzian noise power spectra under CPMG pulse sequences, as we have, which improve the reliability and efficiency of the optimization procedure. On the other hand, Ref.~\cite{Soetbeer2021} requires users to fit the measured $C(t)$ to a stretched exponential, which restricts the form of the power spectrum to $1/f$ noise~\cite{Vezvaee2024}.} In fact, our VQNS offers a systematically improvable means to reconstruct the optimal $S^{\rm trial}(\omega)$ that faithfully reproduces the \textit{experimentally observed} coherence curves, $\{C_j(t) \}$, without assuming \textit{any} noise model or resorting to system-specific preprocessing of data. 

Our design of VQNS facilitates \textit{uncertainty quantification}, \textit{reproducibility}, as well as \textit{systematic improvability} in noise spectrum reconstructions. Because our proposed VQNS algorithm performs a variational optimization on stochastically initialized basis parameters, the results of independent runs converge to power spectra that differ from each other. The extent to which these predictions vary offer a means to quantify the confidence intervals of our predictions, such that narrower confidence intervals indicate the robustness and reproducibility of the power spectrum reconstruction. These confidence intervals primarily depend on three factors: (i) the signal's temporal resolution, (ii) the error bars of the coherence measurements, and (iii) the spectral region of sensitivity of the applied pulse sequences. The temporal resolution, for example, sets an upper limit on the maximum frequency that our reconstruction can resolve. In the limit of low to no experimental noise, our VQNS can accurately reconstruct the noise spectrum in the Fourier-transform allowed region. Even in the Fourier transform-forbidden regions, our VQNS reconstructions remain largely accurate (see our model reconstructions in Fig.~3~(a) and (b) and Fig.~\ref{fig-FTNScomp}~(d) in SI~Sec.~\ref{app:FTNS}) because of its use of a physically motivated Lorentzian basis. However, in the limit of large experimental measurement errors (second factor) which mask as high-frequency oscillations, our VQNS's ability to optimally reconstruct power spectra that faithfully reproduce coherence measurements leads to unphysical high-frequency features. To tame these features, we suggest three options: additional experimental measurements to reduce experimental error, batch subsampling to check for consistency, and mild regularization (see our experimental reconstructions in SI~Sec.~\ref{app:exptMisc}). To address the third factor of spectral sensitivity, we developed and showed the viability of our time-integrated filter function heuristic to determine \textit{which} pulse sequence one should use to improve confidence intervals (see Fig.~5).

\subsection{Advantages over FTNS}
\label{app:FTNS}

Since we have detailed the advantages FTNS offers over DDNS in Ref.~\cite{Vezvaee2024}, we compare the performance of our VQNS approach with that of FTNS~\cite{Vezvaee2024} under equal constraints to highlight its advantages. In particular, we target two known weaknesses of the FTNS methods: the sensitivity to measurement uncertainties in the coherence curves and the temporal resolution of the data. 

Figure~\ref{fig-FTNScomp}(a) shows simulated coherence curves under FID and SE measurements, with an effective 1.0\% artificial measurement noise added to mimic the output of experiments. In Fig.~\ref{fig-FTNScomp}(c), we compare the performances of spectrum reconstruction using the FID FTNS, SE FTNS, and our new VQNS method. To ensure a fair comparison, we used \textit{only} FID and SE coherence measurements as inputs to our VQNS to reconstruct the noise spectrum. As shown in Fig.~\ref{fig-FTNScomp}, the FID and SE coherence curves are simulated up to the time when their (true) values drop to $\sim$0.005 to reflect the fact that, for FTNS, one cannot extract meaningful information beyond the time point in which the coherence hits a value close enough to zero. 
For the FTNS methods, we also apply a minimal denoising procedure. This denoising procedure consists of replacing the signal of $\chi(t)$ beyond $t=4.8555/\omega_0$ with a linear fit of the data between $t=4.8555/\omega_0$ and $t=8.632/\omega_0$ for the FID coherence data, and beyond $t=5.355/\omega_0$ with a linear fit of the data between $t=5.355/\omega_0$ and $t=9.52/\omega_0$ for the SE coherence data (for the justifications for this denoising procedure, see Ref.~\cite{Vezvaee2024}). In contrast, \textit{VQNS does not require denoising procedures of any kind}.

\begin{figure*}[t]
\hspace{-.5cm}{\includegraphics[width=0.8\textwidth]{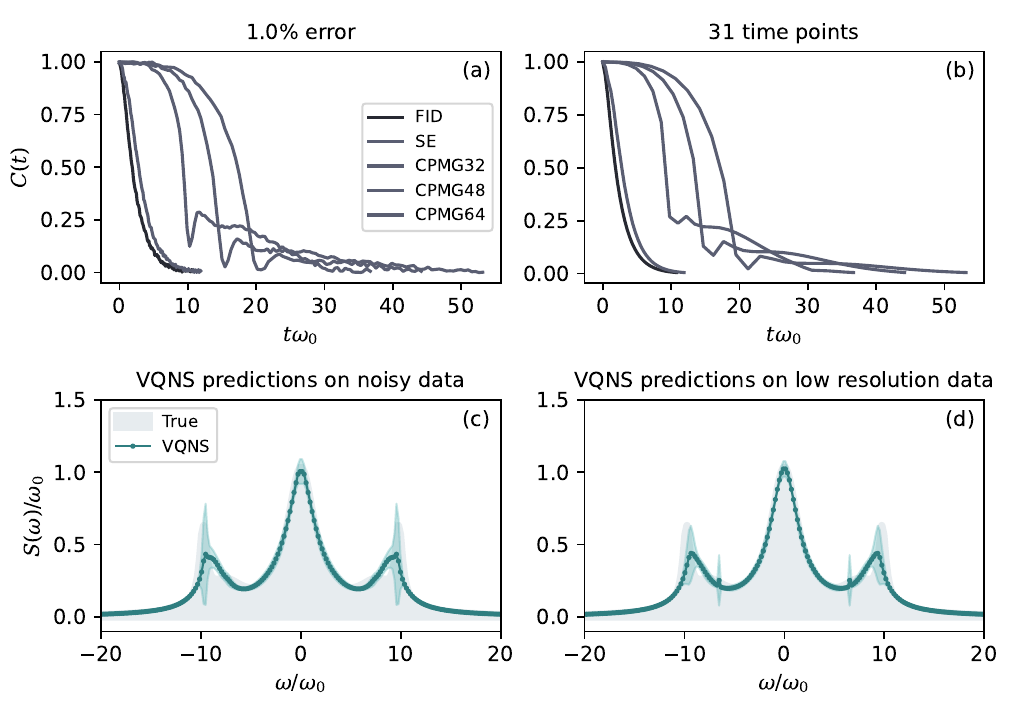}}
\vspace{-10pt}
\caption{(a) Simulated FID, SE, CPMG32, CPMG48, and CPMG64 coherence measurements with an added 1.0\% simulated uncorrelated noise, and (c) VQNS predictions using these as input. For these data, we used a loss threshold of $\xi=8\times10^{-4}$, $N_{\rm basis}=3$, and $N_{\rm runs}=20$. (b) Simulated FID, SE, CPMG32, CPMG48, and CPMG64 coherence measurements with decreased temporal resolution (31 total measurement points) and (d) VQNS predictions using these as input. For these data, we employed a loss threshold of $\xi=7\times10^{-4}$, $N_{\rm basis}=3$, and $N_{\rm runs}=20$.} 
\label{fig-FTNScomp-addCPMG}
\vspace{-10pt}
\end{figure*}   


Statistical error in the input measurement significantly hinders the performance of FTNS while VQNS demonstrates remarkable robustness. The FID FTNS reconstructions in Fig.~\ref{fig-FTNScomp}(c) suffer more in the presence of this effective error, manifesting spurious oscillatory features that become especially prominent at higher frequencies. On the other hand, SE FTNS predicts the structure at intermediate frequencies, $\omega \in [2,8] \omega_0$, relatively well, but struggles to capture the structure of the zero-frequency peak and the high-frequency peaks of the reference spectrum. This observation aligns with the fact that the SE filter function is heavily suppressed at $\omega \rightarrow 0$. In contrast, VQNS offers a more reliable reconstruction of the \textit{entire} spectrum, correctly capturing the peak height and width of the low-frequency contribution and the presence of and approximate position of the high-frequency feature. 

We now compare the performances of the FTNS methods and VQNS for data with limited temporal resolution in the absence of experimental noise. In particular, as a consequence of the Fourier transformation, the temporal resolution of the measured coherence curves limits the maximum resolvable frequency accessible with FTNS. To test this, we reduce the temporal resolution of the coherence curves in Fig.~\ref{fig-FTNScomp}(a) in the absence of added noise from 101 equally distributed time points to 31, keeping the initial and final measurement times fixed (Fig.~\ref{fig-FTNScomp}(b)). 

As before, FTNS performance worsens significantly with poorer resolution of the input signals while VQNS remain largely unchanged. Figure~\ref{fig-FTNScomp}(d) compares the noise spectrum reconstructions obtained from FID FTNS, SE FTNS, and VQNS. Both FTNS methods encounter limits on the maximum resolvable frequency (vertical dotted lines) that render them insensitive to the high-frequency features at $\omega \approx \pm 8$ in the test spectrum. In contrast, since VQNS relies on the variational construction of a trial spectrum from a physical Lorentzian basis, its performance does not suffer from sparse sampling of the coherence curve. At low frequencies, the agreement of the predictions with the true spectrum is equally good for FID FTNS and VQNS, although FID~FTNS is insensitive to the presence of the higher frequency peak and so the reconstructed spectrum becomes inaccurate as the frequency approaches the Fourier limits. SE FTNS again suffers due to the SE filter function being insensitive to the $\omega\rightarrow 0$ limit of the power spectrum. One may explain the underestimation of the high-frequency peak amplitudes across all methods, even in the absence of simulated measurement error, by noting that only the FID and SE coherence curves are used for spectrum reconstruction and their filter functions are most sensitive to low-frequency noise contributions. Despite underestimating the high-frequency peaks, VQNS is the only method that captures them and reproduces the correct asymptotic behavior at large frequencies. 

Another distinct advantage of VQNS over FTNS is that one can improve the former's reliability by adding more measurements under different pulse numbers, whereas FTNS cannot. To demonstrate this point, Figure~\ref{fig-FTNScomp-addCPMG} shows the results of applying VQNS on the data used in Fig.~\ref{fig-FTNScomp}, but with additional CPMG measurements with 32, 48, and 64 pulses. We see that incorporating these pulses improves the prediction of the high-frequency peaks in both the noisy and low-resolution data.

Thus, VQNS is limited neither by maximum frequency constraints nor by the effects of the measurement error in input coherence curves. In addition, the ability of VQNS to achieve this level of efficacy in spectrum prediction {\it without the need for denoising procedures} makes it a practical choice for analyzing experimental data.

\begin{figure}[t]
\hspace{-.5cm}{\includegraphics[width=0.5\columnwidth]{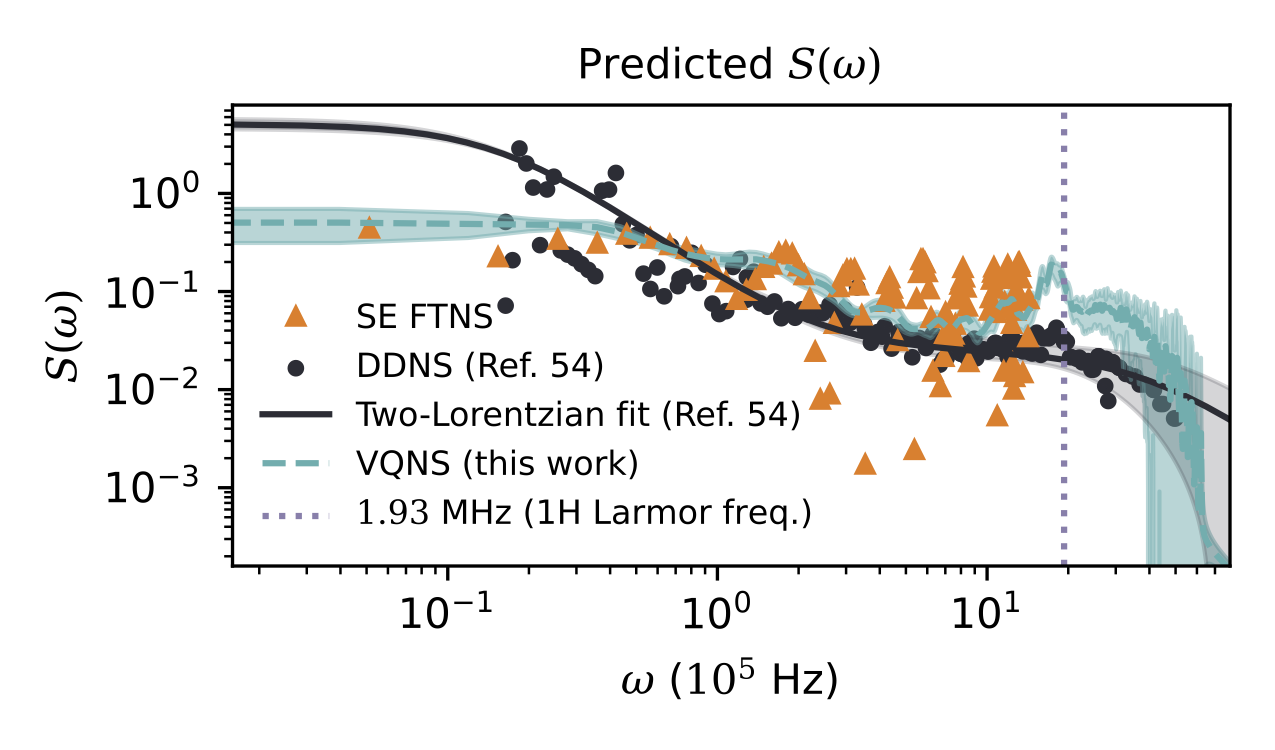}}
\vspace{-15pt}
\caption{Spin echo FTNS reconstruction of the noise spectrum of the surface NV considered in Ref.~\cite{Romach2015}, based on the spin echo experimental measurement. The fact that the measurement exhibits large fluctuations diminishes the efficacy of the FTNS method, leading to a large uncertainty in the reconstructed spectrum.} 
\label{fig-SEFTNS-NV}
\vspace{-20pt}
\end{figure}

As a final demonstration, we compare the performance of applying spin echo FTNS to the experimental data from Ref.~\cite{Romach2015} and compare its reconstructed spectrum against the VQNS predicted spectrum (see Fig.~\ref{fig-SEFTNS-NV}). Although the magnitudes of the FTNS spectrum agree well with the VQNS spectrum, large amplitude oscillatory features start to dominate the structure of the FTNS spectrum toward higher frequencies ($\gtrsim 10^5$ Hz), and it becomes difficult to extract reliable features from the reconstruced spectrum beyond this point. Furthermore, the upper frequency limit up to which FTNS can reconstruct the spectrum does not quite reach the hydrogen Larmor frequency value, making it blind to this important signal. In contrast, our VQNS comfortably captures the hydrogen Larmor signal.

\bibliography{biblo-optNS}